\documentclass[preprint,12pt,authoryear]{elsarticle}

\usepackage{amssymb}
\usepackage{epsfig}

\journal{New Astronomy}

\begin{document}

\begin{frontmatter}

\title{LIGHT CURVE SOLUTIONS OF THE ECCENTRIC BINARIES KIC~10992733, KIC~5632781, KIC~10026136 AND THEIR
OUT-OF-ECLIPSE VARIABILITY}

\author{Diana Kjurkchieva and Doroteya Vasileva}

\address{Department of Physics and Astronomy, Shumen University, 115
Universitetska, 9700 Shumen, Bulgaria}

\begin{abstract}
We determined the orbits and stellar parameters of three eccentric
eclipsing binaries by light curve solutions of their \emph{Kepler}
data. KIC 10992733 and KIC 5632781 undergo total eclipses while
KIC 10026136 reveals partial eclipses. The components of the
targets are G and K stars. KIC 10992733 exhibited variations which
were attributed to variable visibility of spot(s) on
asynchronously rotating component. KIC 5632781 and KIC 1002613
reveal tidally-induced features at periastron, i.e. they might be
considered as eclipsing heartbeat stars. The characteristics of
the periastron features (shape, width and amplitude) confirm the
theoretical predictions.
\end{abstract}


\end{frontmatter}

\section{INTRODUCTION}

The eccentric binaries are important objects for the modern
astrophysics because they present probes for study of tidal
interaction. The gradual loss of energy due to tidal forces leads
to circularization of the orbit and the synchronization of the
rotation of these stars with their orbital motion (Regos et al.
2005).

Recently we are witnesses of the next stage of development of the
tidal interaction theory and testing of its predictions. The
unprecedented high-accuracy \emph{Kepler} observations (Borucki et
al. 2010, Koch et al. 2010) allowed to discover and investigate
new, fine, tidally-induced effects predicted by Kumar et al.
(1995): light feature at the periastron and tidally-excited
oscillations (Welsh et al. 2011, Fuller $\&$ Lai 2011, Burkart et
al. 2012, Thompson et al. 2012, Nicholls $\&$ Wood 2012, Hambleton
et al. 2013a). The newly discovered objects were called
''heartbeat'' (HB) stars. They provide important tests for the
stellar astrophysics and information about the stellar interiors
(Hambleton et al. 2013b).

The number of HB stars found in the \emph{Kepler} data gradually
increases (Kirk et al. 2016). Some of them have been objects of
follow-up spectroscopy which shows a good agreement between the
spectroscopic and photometric orbital elements (Smullen $\&$
Kobulnicky 2015, Shporer et al. 2016, Dimitrov et al. 2017). Last
years we also found 12 HB stars between several samples of
eclipsing \emph{Kepler} systems (Kjurkchieva $\&$ Vasileva 2015,
2016, Kjurkchieva et al. 2016a, b).

The goal of this study is determination of the orbits and physical
parameters of three eccentric binaries, KIC 10992733, KIC 5632781
and KIC 10026136, as well as searching for tidally-excited
effects. Table 1 presents available information for the targets
from the EB catalog (Prsa et al. 2011, Slawson et al. 2011, Kirk
et al. 2016).

\begin{table*}
\caption{Parameters of the targets from the EB catalog: orbital
period \emph{P}, mean temperature $T_{m}$, widths $w_{1,2}$ of the
eclipses (in phase units), depths of the eclipses $d_{1,2}$ (in
flux units), phase of the secondary eclipse $\varphi_{2}$}
\label{tab:1}
\centering
\begin{scriptsize}
\begin{tabular}{cccccccc}
\hline\hline
Kepler ID   &  $P$ [d] & $T_{m}$ [K] & $w_1$    &  $w_2$    & $d_1$     & $d_2$     & $\varphi_{2}$  \\
\hline
10992733    &   18.525  &   5274    &   0.0113  &   0.0153  &   0.405   &   0.268   &   0.72   \\
5632781     &   11.025  &   5786    &   0.04    &   0.0457  &   0.395   &   0.376   &   0.66    \\
10026136    &   9.08    &   6292    &   0.0361  &   0.0532  &   0.3124  &   0.2529  &   0.65   \\
\hline\hline
\end{tabular}
\end{scriptsize}
\end{table*}

\begin{figure*}
   \centering
   \includegraphics[width=13.5cm, angle=0]{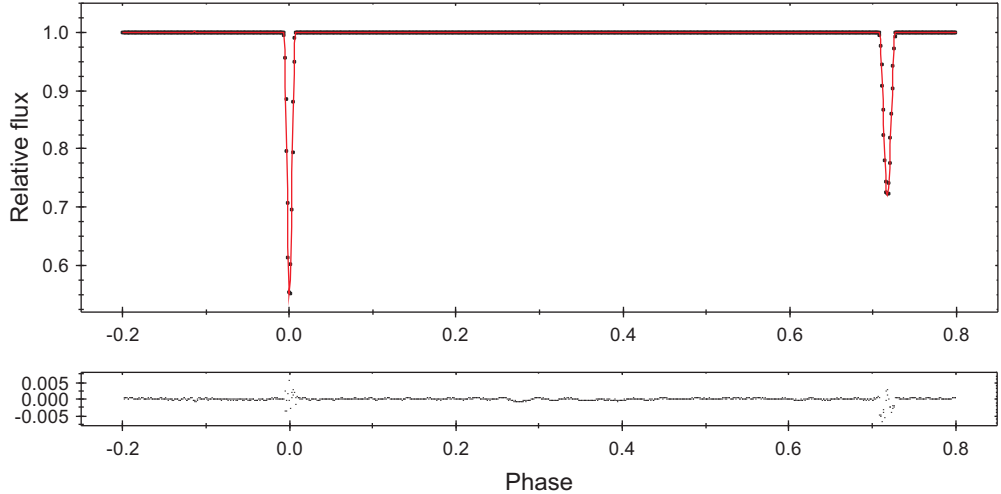}
    \caption{Top panel: the binned folded \emph{Kepler} data of KIC 10992733
    (black points) with their \textsc{PHOEBE} fit (red line);
    Bottom panel: the corresponding residuals from the model subtracted data for all phases (including eclipses)}
   \label{Fig1}
   \end{figure*}

\begin{figure*}
   \centering
   \includegraphics[width=13.5cm, angle=0]{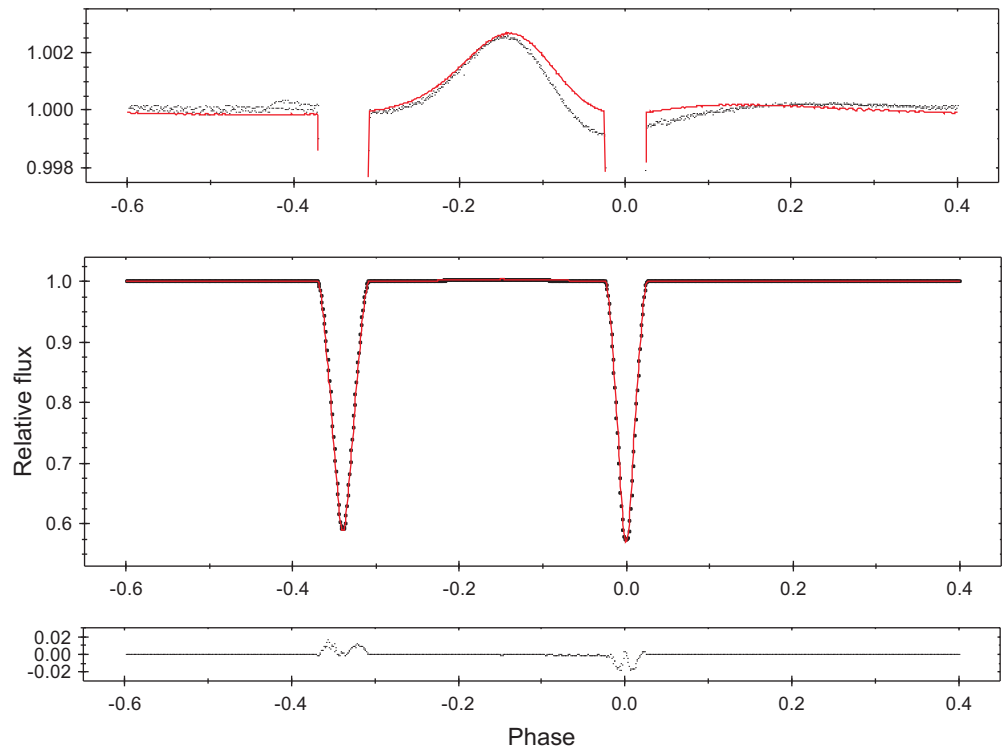}
    \caption{Central panel: the binned folded \emph{Kepler} data of KIC 5632781
    (black points) with their \textsc{PHOEBE} fit (red line);
    Bottom panel: the corresponding residuals from the model subtracted data for all phases (including eclipses);
    Top panel: magnified view of the out-of-eclipse flux variation}
   \label{Fig2}
   \end{figure*}

\begin{figure*}
   \centering
   \includegraphics[width=13.5cm, angle=0]{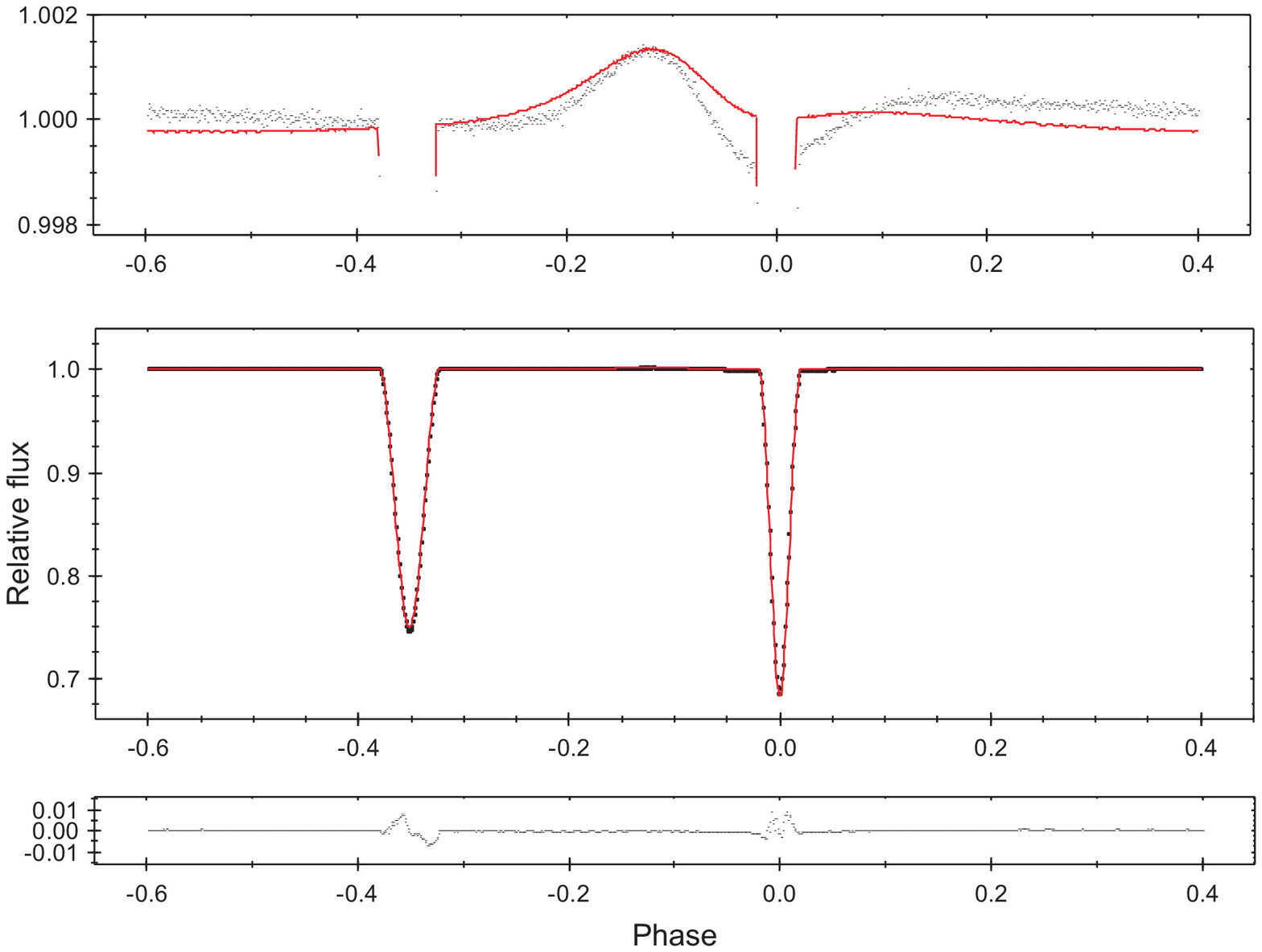}
    \caption{The same as figure 2 for KIC 10026136}
   \label{Fig3}
   \end{figure*}

\section{Light curve solutions}

The modeling of \emph{Kepler} data (Archive MAST) was carried out
by the package \textsc{PHOEBE} (Prsa $\&$ Zwitter 2005). We used
its mode ''Detached binaries'' because our targets are detached
systems, which is clearly true from their light curves (Fig. 1).

Long cadence (LC) data from many quarters are available in the
\emph{Kepler} archive for these binaries. We worked with the live
version of the \emph{Kepler} EB catalog
(http://keplerebs.villanova.edu/). To ignore the effect of
accidental light fluctuations in the procedure of the light curve
solutions and to obtain adequate configuration parameters we
modelled all available LC photometric data (above 50000 points for
each target) after appropriate phase binning. For this aim we
applied the PHOEBE option for binning of data. We used 1000 bins
in phase (the smaller value leads to loss of information from the
eclipses).

The modeling procedure is described in Kjurkchieva $\&$ Vasileva
(2015). Shortly it consists of several steps: fixing $T_1$ to
target temperature $T_m$; preliminary calculation of the
eccentricity \emph{e} and periastron angle $\omega$; varying of
\emph{e} and $\omega$ to fit the eclipse phases; varying of the
inclination \emph{i}, mass ratio \emph{q}, secondary temperature
$T_{2}$ and potentials $\Omega_{1,2}$ to reproduce the whole
curve; adjusting the real stellar temperatures $T_{1}^f$ and
$T_{2}^f$ around the target value $T_{m}$ by the formulae (Ivanov
et al. 2010):

\begin{equation}
T_{1}^f = T_m + s \frac{\Delta T}{s+1},
\end{equation}
\begin{equation}
T_{2}^f = T_1 - \Delta T,
\end{equation}
where $s=l_2/l_1$ (ratio of relative luminosities) and $\Delta T =
T_m-T_2$ were taken from the last \textsc{PHOEBE} fitting.

\begin{table}
\caption{Fitted and calculated parameters of the light curve
solutions: eccentricity $e$; periastron angle $\omega$;
inclination \emph{i}; mass ratio \emph{q}; temperature of the
secondary star $T_{2}$; surface potentials $\Omega_{1, 2}$;
relative radius of the primary and secondary stars $r_{1, 2}$;
periastron phase $\varphi_{per}$; ratio of the relative
luminosities of the stellar components $l_2/l_1$; calculated
temperatures of the primary and secondary stars $T_{1, 2}^f$}
\label{tab:3} \centering
\begin{scriptsize}
\begin{tabular}{cccc}
\hline\hline
Kepler ID       &   10992733    &   5632781 &   10026136    \\
\hline
\emph{e}        &   0.380 (1)   &   0.273(2)&   0.296(1)    \\
$\omega$ [deg]  &   25.34 (1)   &   22.5(1) &   35.95(1)    \\
\emph{i} [deg]  &   89.92 (6)   &   89.2(2) &   87.33(2)    \\
\emph{q}        &   0.509(3)    &   0.97(1) &   0.42(1) \\
$T_2$  [K]      &   4890(24)    &   5739(20)&   6193(20)    \\
$\Omega_1$      &   32.8(2)     &   11.41(4)&   11.1(2) \\
$\Omega_2$      &   23.3(1)     &   13.01(5)&   8.7(1)  \\
\hline
$r_1$           &   0.031(1)    &   0.0993(4)&   0.094(1)    \\
$r_2$           &   0.023(1)    &   0.0834(4)&   0.061(1)    \\
$\varphi_{per}$ &   0.915       &   0.886    &   0.921   \\
$l_{2}/l_{1}$   &   0.471       &   0.685    &   0.392   \\
$T_1$  [K]      &   5386(35)    &   5805(16) &   6319(14)    \\
$T_2$  [K]      &   4995(23)    &   5758(20) &   6220(20)    \\
\hline\hline
\end{tabular}
\end{scriptsize}
\end{table}

Although \textsc{PHOEBE} works with potentials, it gives a
possibility to calculate directly all values (polar, point, side,
and back) of the relative radius $r_i=R_i/a$ of each component
($R_i$ is linear radius and \emph{a} is orbital separation).
Moreover, \textsc{PHOEBE} yields as output parameters bolometric
magnitudes $M_{bol}^i$ of the two components in conditional units
(when radial velocity data are not available). But their
difference $M_{bol}^2-M_{bol}^1$ determines the true luminosity
ratio $s=L_2/L_1=l_2/l_1$.

The method of differential corrections was used as a fitting
algorithm. The parameters of the best light curve solutions are
given in Table 2 while the corresponding synthetic curves are
shown in Figs. 1--3 as continuous lines. The residual curves show
bigger scatters during the eclipse phases (Figs. 1--3). Similar
behavior could be seen also for other \emph{Kepler} binaries
(Hambleton et al. 2013a, Hambleton et al. 2013b, Lehmann et al.
2013, Maceroni et al. 2014). It was attributed to the effects of
finite integration time (Kipping 2010). The reasons for this
effect may be also numerical imperfectness of the physical model
(Prsa et al. 2016) as well as contribution of pulsations and
spots.

The formal PHOEBE errors of the fitted parameters were
unreasonably small. That is why we estimated the parameter errors
manually based on the following rule (Dimitrov et al. 2017). The
error of parameter $b$ corresponded to that deviation $\Delta b$
from its final value $b^{f}$ for which the mean residuals of the
HB features increase by 3$\bar{\sigma}$ ($\bar{\sigma}$ is the
mean photometric error of the target).

The light curve solutions show that: (i) KIC 10992733 and KIC
5632781 undergo total eclipses and their mass ratios should be
considered with a big confidence (Terrell $\&$ Wilson 2005); (ii)
the targets have moderate eccentricity (0.27--0.38); (iii) the
temperatures of target components are close (within 400 K, Table
2); (iv) mass ratios are in the range 0.42--0.97.

\section{Out-of-eclipse variability of the targets}

The out-of-eclipse light of KIC 10992733 contains semi-regular
oscillations (Fig. 4) with periods 5--7 d which undergo long-term
modulation with amplitudes up to 0.005 mag. Their shape as well as
lack of strong periodicity seem to exclude pulsation explanation.
This is confirmed by the periodogram analysis of the
out-of-eclipse data that reveals bad-defined peak (Fig. 5). Its
period $P_{{out}}$ = 5.137 days is not harmonic of the orbital
period (18.525 d). The ratio $P_{orb}/P_{out} \sim$ 3.6 of KIC
10992733 differs almost twice from the ratio of the
pseudo-synchronous angular velocity and mean motion $\Omega_{ps}/n
\sim$ 1.96 for $e$ = 0.38 (Hut 1981). Probably, $P_{{out}}$ =
5.137 days presents rotational period of some of the target
components. Then the observed out-of-eclipse variability may be
explained by variable spot visibility, the bad-defined peak of the
Fourier transform may be result of differential rotation while the
amplitude modulation may be attributed to spot activity cycle. The
components of KIC 10992733 are K stars for which it is inherent
characteristic. The asynchronism of the binary is expected
considering its high eccentricity.

Cool spot with angular size of order of 10$^{\circ}$ may reproduce
the observed amplitude of out-of-eclipse oscillations of KIC
10992733. But there is a problem with the shape of oscillations:
if the system is coplanar one spot of highly-inclined
configuration would cause light variation whose shape is flat
during almost half a cycle, different from the observed one. This
problem may be overcame if one supposes that KIC 10992733 is not
coplanar binary (expected for asynchronous system), i.e. far away
from equilibrium state. Another alternative is KIC 10992733 to be
coplanar binary with two diametrically opposed spots on some of
its components but with a period of 2$P_{{out}}$ = 10.274 d. Then
the ratio $P_{orb}/2P_{out} \sim$ 1.8 almost corresponds to
pseudo-synchronous state of KIC 10992733. Precise spectral
observations in the future may solve this ambiguity.

\begin{figure}
   \centering
   \includegraphics[width=11cm, angle=0]{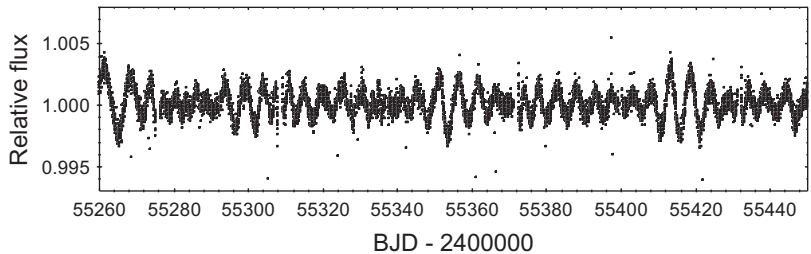}
   \caption{The small-amplitude oscillations of KIC 10992733
   probably are due to photospheric spot.}
   \label{Fig4}
   \end{figure}

   \begin{figure}
   \centering
   \includegraphics[width=7cm, angle=0]{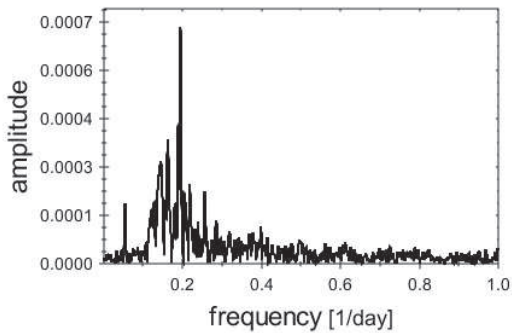}
   \caption{Periodogram of out-of-eclipse variations of KIC 10992733}
   \label{Fig5}
   \end{figure}

The out-of-eclipse light curves of KIC 5632781 and KIC 10026136
reveal light features around the periastron phases (Figs. 2--3,
top panels) with amplitudes of 0.0035 mag and 0.0025 mag, i.e.
these targets are eclipsing heartbeat stars (EB+HB type, Dimitrov
et al. 2017). None of them reveals tidally-induced pulsations
(Figs. 2--4).

The synthetic \textsc{PHOEBE} features at the periastron reproduce
well (but not perfectly) the observed ones. The possibility to
reproduce such mmag details is a compliment of the physical model
of binary stars, particularly of \textsc{PHOEBE}.

The mean temperatures of KIC 5632781 and KIC 10026136 (Table 1)
correspond to the maximum of temperature distribution of the MS
heartbeat stars (figure 2 in Hambleton et al. 2013b).

According to the model of Kumar et al. (1995) the amplitude of the
periastron feature depends on the separation of the objects and
their masses while its shape depends on the orbit parameters
(Hambleton et al. 2013b): the argument of periastron $\omega$
determines the symmetry, the eccentricity \emph{e} determines its
relative width and the inclination \emph{i} dictates the peak to
trough ratio of the tidal pulse (see fig. 5 of Thompson et al.
2012). The contributions of the different parameters on the shape,
width and amplitude of the periastron feature are apparent in
figs. 3--4 of Dimitrov et al. (2017). It should be pointed out
that both model sets (Thompson et al. 2012 and Dimitrov et al.
2017) describe result of clean tidal distortions without
contribution of reflection and other faint effects. The last ones
may be considerable for some cases (as KOI-54).

The tidally-induced periastron features of KIC 5632781 and KIC
10026136 confirm the theoretical predictions.

(a) Their widths (around 0.40--0.45 in phase units) correspond to
eccentricities of around 0.3.

(b) The tidally-induced periastron features have a ''hook'' shape
consisting of two parts with almost the same widths: light
increasing followed by light trough (that coincides with the
primary eclipse). This shape is expected for systems with $0 \leq
\omega \leq 90^0$ (see fig. 3 of Dimitrov et al. 2017). The
arguments of periastron $\omega$ of our two HB targets (Table 2)
fulfil this condition.

(c) The ratios between the amplitudes of the two sections of the
periastron features, the hump-shape part and the trough-shape
part, of our targets are almost 1:1 (Figs. 2--3). This corresponds
to binaries with $\omega \sim 30^0$ (fig. 3 of Dimitrov et al.
2017).

The noneclipsing stars KIC~3547874, KIC 7622059 and KIC~11494130
from the sample of Thompson et al. (2012) exhibit tidally-induced
hook-shape periastron features similar to ours. But their
amplitudes of 1690, 1020 and 600 ppm are considerably smaller than
those of our two targets. This result may be explained by almost 2
times shorter periods of our KIC 5632781 and KIC 10026136 than
those of KIC~3547874, KIC 7622059 and KIC~11494130, because the
shorter period means smaller separation, the stronger tidal
interaction and correspondingly the bigger amplitude of the
tidally-induced periastron feature.

\section{Conclusions}

This paper presents the results of determination of the orbits and
parameters of stellar configurations of the eclipsing eccentric
binaries KIC 10992733, KIC 5632781 and KIC 10026136 on the basis
of their \emph{Kepler} data. The out-of-eclipse light variations
of KIC 10992733 probably due to spot(s). KIC 5632781 and
KIC 10026136 are eclipsing HB stars. The characteristics of
their tidally-induced features are consistent with the theoretical
predictions.

The presented study adds new two members to the family of HB
binaries and provides new data to search for dependencies of the
tidally-induced effects of eccentric binaries on their orbital and
stellar parameters.

\section*{Acknowledgments}
This work was supported partly by project DN08/20 of the Fund for
Scientific Research of the Bulgarian Ministry of Education and
Science and by project RD 08-102/03.02.17 of Shumen University. It
used the SIMBAD database and NASA Astrophysics Data System
Abstract Service. We used data from \emph{Kepler} EB catalog
(http://keplerebs.villanova.edu/). The authors are very grateful
to the anonymous referee for the valuable recommendations and
notes.

\end{document}